\documentclass[reprint,amsmath,amssymb,aps,pra,superscriptaddress,longbibliography]{revtex4-1}

\usepackage{graphicx} 
\usepackage{dcolumn} 
\usepackage{bm} 

\usepackage{braket}
\usepackage{siunitx}
\usepackage{xcolor}
\usepackage{xprintlen}
\definecolor{colBlue}{RGB}{46,48,146}

\usepackage{hyperref}
\hypersetup{
    colorlinks,
    linkcolor=colBlue,
    citecolor=colBlue,
    urlcolor=colBlue
}

\let\oldcite\cite
\renewcommand{\cite}[1]{\mbox{\oldcite{#1}}}

\newcommand{\JKU}{Institute for Theoretical Physics, Johannes Kepler
University Linz, Altenberger Straße 69, 4040 Linz, Austria}
\newcommand{\TUW}{Vienna Center for Quantum Science and Technology,
Atominstitut, TU Wien, Stadionallee 2, 1020 Vienna, Austria}

\begin{document}

\title{Non-Additivity of Microwave-Shielded Interactions between Dipolar Molecules}

\author{Fabian Prieschl}
\affiliation{\JKU{}}
\author{Andreas Schindewolf}
\affiliation{\TUW{}}
\author{Robert E. Zillich}
\affiliation{\JKU{}}

\date{\today}

\begin{abstract}
    Theoretical models of the effective interactions between microwave-shielded polar molecules typically rely on the assumption of pairwise additivity. In this work, we test the validity of this pairwise approximation by evaluating the exact many-body potential energy for configurations of three and four molecules. We demonstrate that this approximation fails for molecule configurations in the plane of the circularly polarized microwave field. In these planar geometries, many-body effects significantly increase the depth of the attractive potential well and shift the position of the repulsive barrier. Although out-of-plane interactions in three-dimensional arrangements weaken this non-additivity, the effective interactions between microwave-shielded dipoles are not a simple sum of pair interactions. In particular, their attractive well, crucial for self-bound clusters, can only be described by the full many-body potential energy.
\end{abstract}

\maketitle

\section{Introduction}

Atoms with large magnetic dipole moments \cite{chomaz_dipolar_2023} introduced a new type of interaction to the field of ultracold quantum gases, leading to a multitude of new phenomena such as droplet phases \cite{chomazPRX16,maciaPRL16,Schmitt2016,Ferrier2016}, or collective excitations previously known only in superfluid helium-4 (rotons) \cite{Santos2003,odellPRL03,hufnaglPRL11,hufnaglPRA13,chomazNatPhys18,Natale2019,Schmidt2021,Blakie2020}.
Owing to their much larger dipole moment, ultracold dipolar molecules \cite{niScience08,deiglmayrPRL08,TakekoshiPRL2014,gregory2026experiments} provide an even richer platform to study novel quantum states with strongly dipolar interaction, such as $p$-wave superfluids \cite{Baranov2002superfluid,Cooper2009Stable,DengSingle}, dipolar quantum Wigner crystals \cite{Buchler2007Strongly,Astrakharchik2007Quantum}, stripe phases \cite{maciaPRL12}, and self-bound membranes \cite{Ciardi2025,Langen2025,Zampronio2026Bilayer}. 

However, inelastic collisions between the molecules inhibited experimental progress in the field for a long time \cite{bause2023ultracold}.
Salvation came with the advent of shielding techniques \cite{Valtolina2020dipolar,Matsuda,anderegg2021observation} that utilize the dipolar interaction to engineer a potential barrier, which repels the molecules at distances much larger than the short range at which the inelastic processes take place.

The most widely used technique is microwave (MW) shielding \cite{Gorshkov2008,Huang2012Field-induced,karman2018,Lassabliere2018Controlling,anderegg2021observation,schindewolf2022evaporation,Schindewolf2026Colloquium}, where a blue-detuned circular MW dresses the rotational transition of the molecules. It has been used to form degenerate Fermi gases of dipolar molecules \cite{schindewolf2022evaporation} and later, with the extension to double MW shielding \cite{DengDouble,Karman2025Double,yuan2025extreme,jozwiak2026map}, enabled Bose--Einstein condensation \cite{Bigagli_2024,shi2026BEC}, the formation of quantum droplets \cite{zhang2026observation}, and Fermi-surface deformation of dipolar molecules \cite{biswas2026controlled}.

The interaction potential of MW-shielded molecules features a long-range potential well, the shape of which is highly tunable through the parameters of the MW field \cite{Schindewolf2026Colloquium}. If this well is deep enough, it can support so-called field-linked bound states \cite{avdeenkov2002collisional,avdeenkov2003linking,Huang2012Field-induced,Lassabliere2018Controlling,Chen2023,chen2024ultracold,Schindewolf2026Colloquium}. Surfacing of these states results in scattering resonances, which can be used to tune the contact interaction, i.e., the $s$-wave \cite{Huang2012Field-induced,Lassabliere2018Controlling,dutta2025universality,DengDouble,Karman2025Double,zhang2026observation,jozwiak2026map,li2026tunable} and $p$-wave interaction \cite{Chen2023,DengSingle,jozwiak2026map}. It has also been demonstrated that coherent formation of field-linked bound states is possible by adiabatically crossing the resonances \cite{chen2024ultracold}. In the presence of the tight optical confinement of an optical lattice, the field-linked states can enable density-induced tunneling resonances \cite{li2026engineering,perez2026hubbard} and tune the on-site interaction \cite{li2026engineering,karman2026low-entropy,stewart-wiese2026hubbard,perez2026hubbard} of extended dipolar Hubbard models \cite{Baranov2012}.

In ultracold atomic systems, two-body collisions can, with little effort, be made intrinsically stable. However, atomic three-body collisions suffer from three-body loss through two-body recombination, which renders the lifetime of atomic few-body bound states, such as Efimov states, to be extremely short \cite{naidon2017efimov,greene2017universal}. Miraculously, MW shielding not only stabilizes the molecules against the inelastic two-body collisions but is also assumed to shield against short-range three-body collisions \cite{yuan2025extreme} and feature fairly stable few-body bound states \cite{Huang2012Field-induced,shi2026universal,wang2026interaction-induced,singh2026efimov}. In the many-body limit, MW-shielded molecules should even be able to form two-dimensional quantum liquids \cite{Langen2025} and quantum crystals \cite{Ciardi2025}.

So far, however, calculations of MW-shielded few- and many-body systems are based on two-body interaction potentials \cite{DengSingle}. The interaction between many molecules is assumed to be a sum of two-body interactions, even at high densities where the mean interparticle distance is comparable to the location of the potential well \cite{Langen2025,Ciardi2025,Zampronio2026Bilayer,shi2026universal,wang2026interaction-induced,singh2026efimov}. For neutral atoms and unshielded molecules this is usually a good approximation, but even there three-body corrections \cite{axilrodJCP43} can be important \cite{cazorlaPRB15}. For microwave-driven polar molecules in optical lattices, Büchler \textit{et al.\ }used a perturbative expansion to derive effective three-body interactions \cite{buchler2007threebody}. Because their approach is restricted to large intermolecular distances, they concluded that four-body and higher-order terms can be safely ignored. However, for MW-shielded molecules in the non-perturbative short-range regime of the attractive well, the validity of approximating the many-body potential as a simple sum of two-body interactions needs to be checked.

In this paper, we demonstrate for single-MW shielding that this approximation indeed fails in configurations that are most vital for few-body bound states, self-bound liquids, and quantum crystals. For clusters of more than three molecules, we show that it is necessary to consider at least four-body interaction potentials in order to provide accurate calculations of resonance positions, binding energies, and lifetimes of few-body field-linked states, elastic and inelastic few-body collision rates, as well as quantum phase transitions of many-body systems.

\section{Theory}
\label{sec:theory}

We want to calculate the energy of $N$ dipolar molecules in a circular near-resonant MW field for a given
arrangement of the relative molecule positions. With their positions fixed, the Hamiltonian $H$ contains rotational, but no translational
degrees of freedom:
\begin{align}
  H = \sum_i H_{\rm rot}^{(i)} + \sum_i H_{\rm MW}^{(i)}(t) + \sum_{i<j} V_{dd}(\mathbf{r}_{i}-\mathbf{r}_{j}).
  \label{eq:H}
\end{align}
$H_{\rm rot}^{(i)}=BL_i^2$ is the free-linear-rotor Hamiltonian of molecule $i$ with rotational constant $B$, and
$H_{\rm MW}^{(i)}(t)=-\hat{\mathbf{d}}^{(i)}\cdot\mathbf{E}(t)$ is the interaction between the dipole $\hat{\mathbf{d}}^{(i)}$ of molecule $i$
and a harmonic MW field $\mathbf{E}(t)$ with angular frequency $\omega$. 
The coupling between molecules is accomplished with the dipole-dipole interaction:
\begin{align}
V_{dd}(\mathbf{r}) = \frac{1}{4\pi\varepsilon_0 r^3} \left[ \hat{\mathbf{d}}^{(i)}\! \cdot \hat{\mathbf{d}}^{(j)}
- 3(\hat{\mathbf{d}}^{(i)}\! \cdot \hat{\bf r})(\hat{\mathbf{d}}^{(j)}\! \cdot \hat{\bf r}) \right],
\label{eq:vdd}
\end{align}
where $\mathbf{r}=\mathbf{r}_i-\mathbf{r}_j$ is the distance vector between molecule $i$ and $j$.
Following the Born-Oppenheimer approximation, the translational motion is decoupled from the rotation of the
dipoles, and the energy associated with $H$ is effectively the potential energy surface (PES) of
the molecules. Here, we are only concerned with the calculation of the many-body PES, not solving the translational
Hamiltonian for scattering and bound states.

For the interaction between two MW-dressed molecules, the orientation of the field is essential for the shape of the effective long-range potential. For a circularly polarized MW field ($\sigma^+$) propagating along the $z$-axis, the electric field vector rotates within the $xy$-plane \cite{karman2018,Lassabliere2018Controlling}. Consequently, the induced dipole moments of the molecules continuously rotate within this plane. When molecules are arranged spatially within the $xy$-plane, the time-averaged dipole-dipole interaction is dominated by configurations where the dipoles point head-to-tail towards each other, yielding an attractive potential. In contrast, the rotating field exhibits cylindrical symmetry around the $z$-axis, resulting in physically identical $xz$ and $yz$ planes. In these planes, the out-of-plane separation ensures the rotating dipoles interact side-by-side, creating a repulsive potential. Our results will show that this qualitative behavior still holds true for more than two molecules, but that quantitatively, the attractive well is a many-body effect that cannot be understood as the sum of two-body interactions.

\subsection{Dressed states of single molecule}
\label{ssec:dressed}

The MW is nearly resonant with the $J=0\to 1$ transition of the molecules. Therefore, we only consider the
rotational subspace consisting of the rotational ground state $J=0$ and the triply degenerate first excited rotational states $J=1$.
Our truncated free-rotor basis is thus $\{|0,0\rangle, |1,1\rangle, |1,0\rangle, |1,-1\rangle\}$, which are spherical harmonics, $Y_{J,M}(\theta, \varphi)$,
in angular coordinate representation.
The free rotational energy is given by $E_J = B J(J+1)$, where $B$ is the rotational constant. In the free-rotor basis, the free-rotor Hamiltonian is diagonal:
\begin{equation}
	H_{\text{rot}} = \hbar \begin{pmatrix} 0 & 0 & 0 & 0 \\ 0 & \omega_0 & 0 & 0 \\ 0 & 0 & \omega_0 & 0 \\ 0 & 0 & 0 & \omega_0 \end{pmatrix},
\end{equation}
where $\hbar\omega_0 = 2B$.

We define the dipole moment operator in the spherical tensor basis as
\begin{equation}
	\hat{d}_q = d \sqrt{\frac{4\pi}{3}} Y_{1,q}(\theta, \varphi),
    \label{eq:dYlm}
\end{equation}
where $d$ is the magnitude of the permanent dipole moment. We can evaluate transition matrix elements using the orthogonality of $Y_{J,M}$:
\begin{align}
	\langle 1, q' | \hat{d}_q | 0,0 \rangle &= \frac{d}{\sqrt{3}} \delta_{q',q}.
\end{align}
The applied MW field is expressed with a complex polarization vector $\vec{\varepsilon}$ in the spherical basis, normalized such that $|\varepsilon_1|^2 + |\varepsilon_0|^2 + |\varepsilon_{-1}|^2 = 1$:
\begin{equation}
	\mathbf{E}(t) = \frac{E_0}{2} \left( \sum_{q \in \{1,0,-1\}} \varepsilon_q \hat{e}_q^* \right) e^{-i\omega t} + \text{c.c.},
\end{equation}
where the spherical unit vectors matching the $Y_{1,q}$ basis are defined as $\hat{e}_1 = -\frac{1}{\sqrt{2}}(\hat{x}+i\hat{y})$, $\hat{e}_0 = \hat{z}$, and $\hat{e}_{-1} = \frac{1}{\sqrt{2}}(\hat{x}-i\hat{y})$. 

The MW Hamiltonian is $H_{\text{MW}}(t) = -\hat{\mathbf{d}} \cdot \mathbf{E}(t) = -\sum_q \hat{d}_q \varepsilon_q^*$, which yields the matrix elements:
\begin{equation}
	\langle 1, q | H_{\text{MW}}(t) | 0,0 \rangle = -\frac{d E_0}{2\sqrt{3}} \varepsilon_q^* e^{-i\omega t} + \text{H.c.}
\end{equation}
We define the Rabi frequency $\Omega = \frac{d E_0}{\hbar\sqrt{3}}$ and its polarization components $\Omega_q = \frac{\Omega}{2} \varepsilon_q$. The total time-dependent internal Hamiltonian in the laboratory frame is simply $H_{\text{lab}}(t) = H_{\text{rot}} + H_{\text{MW}}(t)$.


We apply a unitary transformation to the rotating frame,
\begin{equation}
	U_{\text{rot}}(t) = |0,0\rangle\langle0,0| + e^{-i\omega t}\sum_q |1,q\rangle\langle 1,q|.
\end{equation}
Close to resonance, where the detuning $\delta=\omega-\omega_0\ll\omega_0$ is small, we can employ the rotating wave approximation (RWA), where we discard the highly
oscillatory terms $\sim e^{\pm 2i\omega t}$ to obtain the effective time-independent internal Hamiltonian $H_0$:
\begin{equation}
	H_0 = -\hbar \begin{pmatrix} 0 & \Omega_1 & \Omega_0 & \Omega_{-1} \\ \Omega_1^* & \delta & 0 & 0 \\ \Omega_0^* & 0 & \delta & 0 \\ \Omega_{-1}^* & 0 & 0 & \delta \end{pmatrix}.
\end{equation}
The MW-dressed states are the eigenstates of $H_0$. Using the fact that the polarization vector is normalized
($|\varepsilon_1|^2 + |\varepsilon_0|^2 + |\varepsilon_{-1}|^2 = 1$), we find the eigenfrequencies of two dark states $|D_1\rangle$ and $|D_2\rangle$ and two
bright states $|+\rangle$ and $|-\rangle$,
\begin{align}
    \lambda_{D_1, D_2} &= -\delta,\\
	\lambda_\pm &= \frac{-\delta \pm \sqrt{\delta^2 + \Omega^2}}{2} \equiv \frac{-\delta \pm \Omega_{\text{eff}}}{2},
\end{align}
where we defined $\Omega_{\text{eff}}=\sqrt{\delta^2 + \Omega^2}$. The bright eigenstates are
\begin{equation}
	|\pm\rangle = \alpha_\pm \begin{pmatrix} -(\delta + \lambda_\pm) \\ \Omega_1^* \\ \Omega_0^* \\ \Omega_{-1}^* \end{pmatrix},
\end{equation}
where $\alpha_\pm = [(\delta+\lambda_\pm)^2 + \frac{\Omega^2}{4}]^{-1/2}$.
The dark states are superpositions of only $J=1$ rotational levels that are orthogonal to the polarization vector of the MW field:
$|D_1\rangle=(0, D_{1,1}, D_{1,0}, D_{1,-1})^T$ such that $\Omega_1 D_{1,1}+\Omega_0 D_{1,0}+\Omega_{-1}D_{1,-1}=0$
and similarly for $|D_2\rangle$.
The four eigenvectors $\{|+\rangle, |D_1\rangle, |D_2\rangle, |-\rangle\}$ define the single-molecule
unitary transformation matrix $U_d$ from the eigenstates of $H_{\rm rot}$ to the dressed eigenstates, leading to the diagonalized
single-molecule internal Hamiltonian, $H_{\text{int}} = U_d^\dagger H_0 U_d$.

\subsection{Exchange Symmetry and Subspaces}

Assuming the molecules are initially prepared in identical internal states, the two-molecule internal wavefunction is symmetric under particle exchange. Because the total Hamiltonian preserves this exchange symmetry, the relevant adiabatic PESs are strictly confined to the symmetric subspace. We define the symmetrized (and antisymmetrized) basis states as
\begin{equation}
	|\Psi_{ij}^\pm\rangle = \frac{1}{\sqrt{2(1+\delta_{ij})}} \Big( |i\rangle \otimes |j\rangle \pm |j\rangle \otimes |i\rangle \Big),
\end{equation}
where $i,j \in \{+, D_1, D_2, -\}$.
The 16-dimensional product space decomposes into a 10-dimensional symmetric subspace (4 identical pairs like $|+\rangle\otimes|+\rangle$, and 6 symmetric superpositions) and a 6-dimensional antisymmetric subspace. To improve computational efficiency, states that are physically irrelevant to the shielding barrier can be truncated. Specifically, states constructed entirely from the uncoupled $|D_1\rangle$ and $|D_2\rangle$ bare states (e.g., $|D_1, D_1\rangle$ or $|D_1, D_2\rangle_s$) possess zero ground-state $|0,0\rangle$ character. They remain dark to the MW field and do not participate in the resonant shielding mechanism. By safely discarding these ``double dark'' states, the 10-dimensional space is reduced to a 7-dimensional effective symmetric subspace.

\subsection{The dipole-dipole interaction}

It is mathematically advantageous to recast the dipole-dipole interaction into the basis of irreducible spherical tensors (see Appendix \ref{app:spherical_tensor}). The spatial dependence is absorbed into the Racah-normalized spherical harmonics $C_{2,-p}(\theta_r, \varphi_r) = \sqrt{4\pi/5} Y_{2,-p}(\theta_r, \varphi_r)$, yielding the purely spatial-angular separated form:
\begin{equation}
	V_{dd}(\mathbf{r}) = -\frac{\sqrt{6}}{4\pi\varepsilon_0 r^3} \sum_{p=-2}^{2} (-1)^p C_{2,-p}(\theta_r, \varphi_r) \left[ \hat{d}^{(i)} \otimes \hat{d}^{(j)} \right]_p,
    \label{eq:vddspherical}
\end{equation}
where the rank-2 dipole tensor is defined via Clebsch-Gordan coefficients as 
\begin{equation}
    \left[ \hat{d}^{(i)} \otimes \hat{d}^{(j)} \right]_p = \sum_{q_1, q_2} \langle 1, q_1; 1, q_2 | 2, p \rangle \hat{d}_{q_1}^{(i)} \hat{d}_{q_2}^{(j)}.
\end{equation}

We first apply the two-body rotating frame transformation $V_{dd}^{\text{rot}} = U_{\text{rot}}^{(i,j)\dagger} V_{dd} U_{\text{rot}}^{(i,j)}$ and then implement the rotating wave approximation (RWA) to average out the highly oscillatory phases. As detailed in Appendix \ref{app:rwa_derivation}, this mathematically drops the rapidly oscillating pair-excitation and de-excitation processes, enforcing the conservation of rotational excitations ($\Delta J_i + \Delta J_j = 0$). The resulting time-independent interaction in the RWA is
\begin{equation}
    \begin{split}
        V_{dd}^{\text{RWA}} &= -\frac{\sqrt{6} d^2}{12\pi\varepsilon_0 r^3} \sum_{p=-2}^{2} (-1)^p C_{2,-p}(\theta_r, \varphi_r) \\ &\quad \sum_{q_1, q_2} \langle 1, q_1; 1, q_2 | 2, p \rangle \left( \hat{d}_{q_1}^{+(i)} \hat{d}_{q_2}^{-(j)} + \hat{d}_{q_1}^{-(i)} \hat{d}_{q_2}^{+(j)} \right),
    \end{split}
\end{equation}
where $d$ is the permanent dipole moment, and $\hat{d}_q^{+(i)} = |1,q\rangle_i\langle 0,0|_i$ and $\hat{d}_q^{-(i)} = (-1)^q |0,0\rangle_i\langle 1,-q|_i$ are the dimensionless rotational raising and lowering operators for molecule $i$.

Finally, the RWA-approximated interaction is transformed into the dressed basis via
\begin{equation}
    V_{\text{dressed}}^{(1,2)} = U_d^{(1,2)\dagger} V_{dd}^{\text{RWA}} U_d^{(1,2)},
\end{equation}
where $U_d^{(1,2)} = U_d^{(1)} \otimes U_d^{(2)}$ is the direct product of the two single-molecule unitary transformations calculated in Sec.~\ref{ssec:dressed}. We refrain from providing lengthy analytical expressions for $V_{\text{dressed}}^{(1,2)}$, as this final transformation is performed numerically.

\subsection{$N$-body problem}

So far, our derivation is applicable to two dipoles, as has been previously derived by others \cite{DengSingle}.
To consider $N$ molecules, we transform the rotational $N$-body Hamiltonian Eq.~(\ref{eq:H}) to the dressed basis and then diagonalize it.

The total Hamiltonian $H_{\text{total}}^{(1,\dots ,N)}$ consists of the non-interacting internal Hamiltonians of $N$ molecules, derived in the dressed basis
in Sec.~\ref{ssec:dressed}, and the pairwise dipole-dipole interactions between them:
\begin{equation}
	H_{\text{total}}^{(1,\dots ,N)} = H_{\text{int}}^{(1,\dots ,N)} + V_{\text{dd}}^{(1,\dots ,N)}.
    \label{eq:Hfinal}
\end{equation}
The internal Hamiltonian $H_{\text{int}}^{(1,\dots ,N)}$ is simply the sum of the single-particle terms, see Eq.~(\ref{eq:H}). To be mathematically precise, the
internal Hamiltonian $H_{\rm int}^{(i)}$ of molecule $i$ acts in the subspace of molecule $i$. For proper notation, we embed $H_{\rm int}^{(i)}$
in the many-body Hilbert space of all molecule rotations using the tensor product with the identity matrix $\mathbb{I}$ for all other molecules:
\begin{equation}
	H_{\text{int}}^{(1,\dots ,N)} = \sum_{i=1}^N \left( \mathbb{I}^{\otimes i-1} \otimes H_{\text{int}}^{(i)} \otimes \mathbb{I}^{\otimes N-i} \right).
\end{equation}
Similarly, the total interaction $V_{\text{dd}}^{(1,\dots ,N)}$ is the sum of all pairwise interactions. We can calculate the interaction for a single reference pair
(e.g., molecules 1 and 2). To apply this specific interaction to any arbitrary pair $(i,j)$ in the full Hilbert space, we simply swap indices
with operators $P_{ab}$ which swap the internal states of molecules $a$ and $b$, apply the reference interaction, and swap them back:
\begin{equation}
	V_{\text{dd}}^{(1,\dots ,N)} = \sum_{i<j}^N (P_{1i} P_{2j}) \left( V_{\text{dressed}}^{(1,2)} \otimes \mathbb{I}^{\otimes N-2} \right) (P_{2j} P_{1i}).
\end{equation}
This approach gives us the complete matrix representation of the $N$-body system. Numerical diagonalization of $H_{\text{total}}^{(1,\dots ,N)}$ yields
eigenenergies which are the adiabatic PESs $V_i({\bf r}_1,\dots,{\bf r}_N)$ for $N$ molecules without relying on the pairwise approximation.
We still make the usual assumption that the Born-Oppenheimer approximation is valid, i.e., rotational and translational motion are
decoupled. The quantum number $i$ labels the $4^N$ surfaces; we will focus on the PES that connects for large separations to the highest dressed states of all molecules since this is the experimentally most relevant state (i.e., the shielded state).

\subsection{Many-Body expansion of the PES}

The many-body expansion of the interaction energy \cite{HankinsMBE} allows us to systematically study the contributions of two-body, three-body, and higher-order interactions by isolating irreducible $N$-body effects.
The PES of an $N$ particle system with coordinates ${\bf r}_1,\dots,{\bf r}_N$ can be expanded as a convergent series of cluster terms of increasing order:
\begin{equation}
    \begin{split}
        V({\bf r}_1,\dots,{\bf r}_N) &= \sum_{i} E^{(i)} + \sum_{i<j} V^{(i,j)}({\bf r}_i,{\bf r}_j) \\ &\quad + \sum_{i<j<k} V^{(i,j,k)}({\bf r}_i,{\bf r}_j,{\bf r}_k) + \dots \, .
    \end{split}
	\label{eq:mbe}
\end{equation}
$E^{(i)}$ represents the single-particle energies; they do not depend on the particle position, therefore
we subtract them in all of the figures. The higher-order terms $V^{(i_1,\dots ,i_n)}$ represent the
irreducible $n$-body interactions, the energy correction that cannot be explained by the sum of all lower-order terms.

While the expansion is general, the behavior of the terms $V^{(i_1,\dots ,i_n)}$ depends on the physical system. 
In many cooperative systems, such as lithium clusters \cite{lithium_clusters},
the expansion exhibits an alternating sign structure: if the two-body interaction $V^{(i,j)}$ is repulsive,
the three-body term $V^{(i,j,k)}$ tends to be attractive, and so on.
In bulk water, on the other hand, the many-body terms do not exhibit a simple alternating sign behavior,
leading to slow convergence
where higher-order cooperative effects remain significant \cite{BulkWaterSimulations, DemerdashMBE}.

The second term in Eq.~(\ref{eq:mbe}) constitutes the pairwise approximation
to the full PES, which is exact only for $N=2$, but is often used in molecular simulations.
We will see below that it fails for MW-shielded molecules. For $N=3$, we compare the full
PES with the $V({\bf r}_1,{\bf r}_2,{\bf r}_3)$ pairwise approximation for a few configurations.
For $N=4$, we additionally compare $V({\bf r}_1,{\bf r}_2,{\bf r}_3,{\bf r}_4)$ with the triplet approximation. 
The quadruplet contains four unique triplet sub-clusters ($(1,2,3), (2,3,4), (1,3,4), (1,2,4)$). 
Summing the energies of these four triplets introduces a double-counting error because every connection is shared by exactly two triplets and is thus counted twice.
To correct for this, we need to subtract the sum of the pairwise interactions:
\begin{equation}
    V_{\text{tri}}({\bf r}_1,{\bf r}_2,{\bf r}_3,{\bf r}_4) \approx \sum_{i<j<k} V({\bf r}_i,{\bf r}_j,{\bf r}_k) - \sum_{i<j} V({\bf r}_i,{\bf r}_j).
\end{equation}
The triplet approximation is expected to be an improvement over the pairwise approximation, but we will show that it still differs significantly from the full 4-body potential for attractive configurations.

\begin{figure*}
    \includegraphics[width=0.9\textwidth]{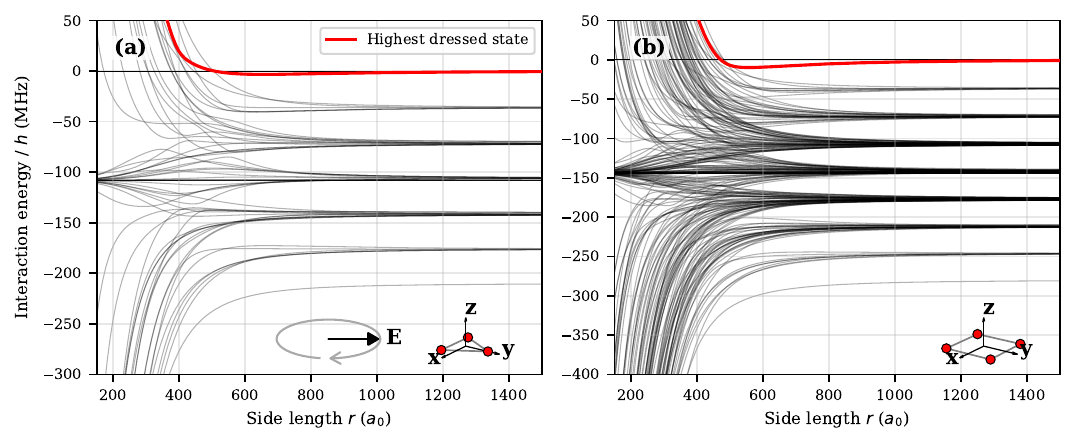}
    \caption{\label{FIG:all} 
    Overview of all adiabatic PESs in the $xy$-plane for 3 dipoles forming an equilateral triangle of side length $r$ (a)
    and of 4 dipoles forming a square of side length $r$ (b). The highest-energy adiabatic surface is highlighted in red. The insets depict the respective molecular configurations in the coordinate frame. Panel (a) additionally illustrates the microwave electric field vector $\mathbf{E}$ rotating in the $xy$-plane.}
\end{figure*}

\section{Results}

We evaluate the PESs of three and four MW-shielded molecules beyond pairwise additivity by matrix diagonalization of $H_{\text{total}}^{(1,\dots ,N)}$, given in Eq.~(\ref{eq:Hfinal}). We stress that our analysis is universally valid for ${}^1\Sigma^+$ molecules, such as $^{23}\text{Na}^{39}\text{K}$ molecules ($d = 2.72$~Debye, $B = h \times 2.848\,$GHz). Unless stated otherwise, calculations assume a blue-detuned MW field with $\Omega = 2\pi \times 70$\,MHz and $\delta = 2\pi \times 2$\,MHz. Although varying these field parameters quantitatively modifies the shape and depth of the potential wells, the qualitative failure of lower-order approximations persists across experimentally accessible regimes. We assume circular polarization of the MW field which propagates in the $z$-direction, such that the electric field vector $\mathbf{E}(t)$ rotates in the $xy$-plane (see schematic inset in Fig.~\ref{FIG:all}).

In Fig.~\ref{FIG:all}, we show all PESs as function of the side length $r$ for the three-body equilateral triangle (a) and the four-body square (b), both lying in the $xy$-plane.
In the blue-detuned regime, the molecules are prepared in the highest dressed state, which is the asymptotic value of the highest-energy adiabatic surface (highlighted in red). As the number of molecules increases from three to four, the spectrum becomes substantially denser, the attractive well deepens, and the crossing point with lower-lying channels shifts to lower interaction energies.

We calculate the PESs for a few important geometric arrangements of the molecules. Indeed, the planar equilateral triangle and square configurations in the $xy$-plane are the global energy minima for the three- and four-body systems, respectively. To confirm that these are global rather than local minima, we performed unconstrained 3D geometric energy minimizations using the Nelder-Mead algorithm \cite{nelder1965} starting from a wide range of randomized 3D initial configurations. In all cases, the optimization converged to the symmetric planar structures, yielding minimum-energy side lengths of $r = 654.34\,a_0$ for the triangle and $r = 556.93\,a_0$ for the square. The triangular configuration is particularly interesting, as its energy minimum determines the geometry of three-body bound states \cite{Huang2012Field-induced}. Beyond that, planar configurations play an essential role in crystallized self-bound membranes \cite{Ciardi2025}. A correct description of the attractive part of the interaction of MW-shielded molecules is crucial for quantitative theoretical predictions. We also investigate linear chains which are relevant for molecules in one-dimensional optical lattices, used to create isolated or coupled tubes \cite{moritz2003,paredes2004,haller2009}. In one-dimensional systems, correlations play an increased role; in combination with the long range of dipole interactions, one-dimensional dipolar gases become an interesting model system, see e.g.\ Ref.~\cite{Baranov2012}. To contrast these planar and linear structures with a fully three-dimensional configuration, we finally consider a regular tetrahedron.
With these geometries, we explore the attractive and repulsive regions, and combinations thereof, discussed qualitatively already at the beginning of Sec.~\ref{sec:theory}, and we will identify the configurations where the many-body nature of the effective interaction is most pronounced.

\subsection{Planar configurations}

\begin{figure}
    \includegraphics[width=0.9\columnwidth]{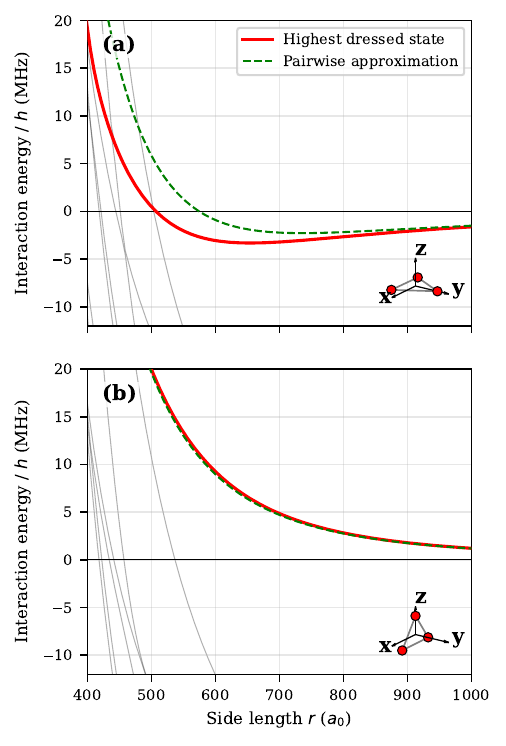}
    \caption{\label{FIG:triangle} 
    Interaction energy of three dipoles arranged in an equilateral triangle in the $xy$-plane (a) and $xz$-plane (b). The solid red curve represents the exact highest-energy dressed state, while the dashed green curve shows the pairwise approximation.
    }
\end{figure}

\begin{figure}
    \includegraphics[width=0.9\columnwidth]{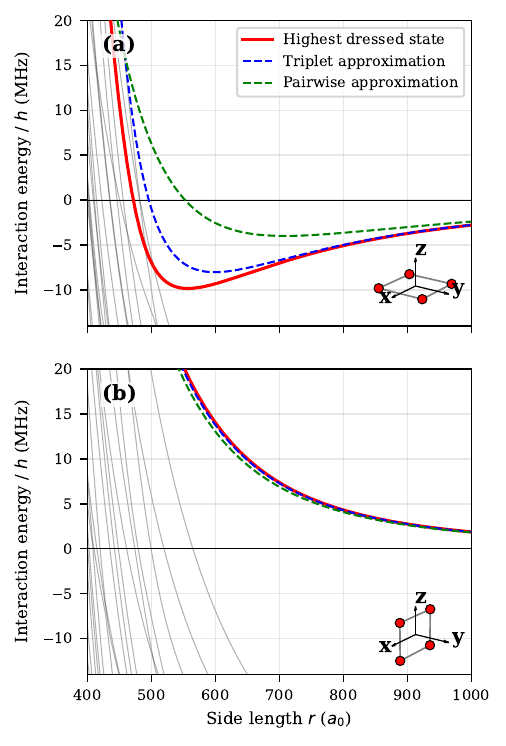}
    \caption{\label{FIG:square} 
    Interaction energy of four dipoles arranged in a square in the $xy$-plane (a) and $xz$-plane (b). The solid red curve corresponds to the exact highest-energy dressed state. The dashed green and blue curves represent the pairwise and triplet approximations, respectively.
    }
\end{figure}

To quantify the non-additivity in these systems, we compare the full many-body interaction energy with lower-order pairwise and triplet approximations. Figure~\ref{FIG:triangle} shows the many-body (here triplet) interaction (red) and pairwise approximation (dashed) for an equilateral triangle. In the $xy$-plane (a), the exact interaction energy exhibits an attractive region before turning repulsive at short range. The pairwise approximation predicts this behavior qualitatively, but it underestimates the strength of the attraction and shifts the repulsive part of the potential to larger intermolecular distances. When the triangle is rotated into the $xz$-plane (b), the breaking of planar symmetry turns exact crossings into avoided crossings (outside the shown range). Out-of-plane repulsion dominates, and both the exact state and the pairwise approximation maintain a purely repulsive potential. Unlike in the attractive configuration in the $xy$-plane, the pairwise approximation is completely adequate for the triangle in the $xz$-plane.  We will see this behavior in all cases: the interaction in the MW-polarization plane is a many-body effect, while it is quite well-described by pair potentials in the perpendicular direction.

The deviation from the pairwise approximation is amplified in the four-body square geometry, shown in Fig.~\ref{FIG:square}. We show the many-body (now quadruplet) interaction (red), the triplet approximation (blue dashed) and the pairwise approximation (green dashed). In the $xy$-plane (a), the pairwise interaction significantly differs from the full interaction, with an energy minimum deviating by almost a factor of three. The triplet approximation provides an improvement over the pairwise sum, but still underestimates the magnitude of the attraction and overestimates the repulsion, indicating that exact four-body interactions are necessary to accurately describe the short-range behavior. However, the improvement from pairwise to triplet potential indicates a convergence towards the final PES. Similar to the triangle, orienting the square in the $xz$-plane (b) yields a purely repulsive potential, where the approximate lower-order energies follow the full many-body energy quite closely.

In all cases, for very large side lengths $r>1000 a_0$, the approximate energies converge to the full interaction, as expected. We have seen the same behavior if we shift one molecule of the triangle to progressively larger distance (not shown). Hence for a low-density BEC of molecules, the pairwise approximation may be sufficient if the probability for three molecules coming close is small enough, and they therefore do not encounter the 3-body potential minimum. However, we stress again that high-density molecule clusters are self-bound precisely because of the attractive well. For these interesting phases of matter, the pairwise approximation fails, because the mean interparticle spacing is of the order of the attractive well \cite{Langen2025}.

\subsection{Linear configurations}

\begin{figure}
    \includegraphics[width=0.9\columnwidth]{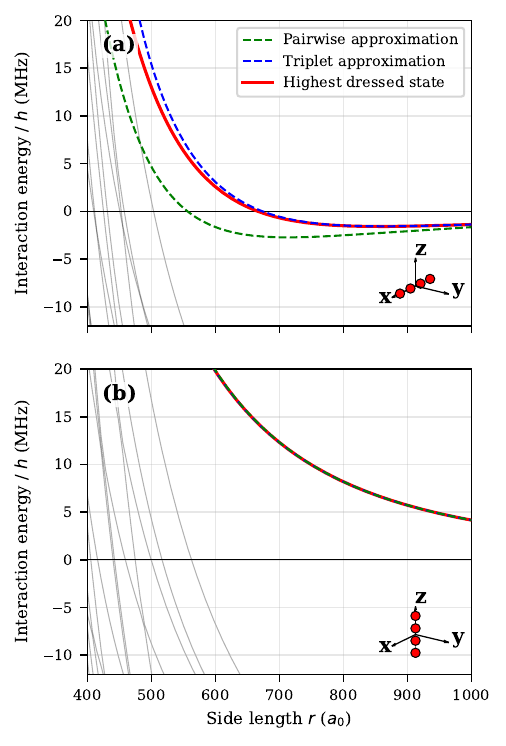}
    \caption{\label{FIG:linear_chain} 
    Interaction energy of four dipoles arranged in a linear chain along the $x$-axis (a) and $z$-axis (b). The solid red curve represents the exact highest-energy dressed state, while the dashed green and blue curves show the pairwise and triplet approximations, respectively.
    }
\end{figure}

To model molecules confined in one-dimensional geometries, e.g., by means of an optical lattice, we evaluated the interaction energy of a four-body linear chain. Figure \ref{FIG:linear_chain} shows the interaction energy of such a chain with equidistant separation $r$. The orientation of the chain relative to the MW polarization determines the behavior of the potential. For a chain aligned along the $x$-axis (i.e., in the MW-polarization plane), the dipoles interact predominantly attractively. In this configuration (a), the exact interaction energy exhibits a shallow attractive well before turning repulsive. In contrast to the square configuration, the pairwise approximation overestimates the magnitude of this attraction for the linear chain and shifts the repulsive part of the potential to much shorter intermolecular distances. Conversely, the triplet approximation slightly overestimates the repulsive many-body contributions, but approximates the attraction very well.

Aligning the chain along the $z$-axis ensures that the rotating dipoles interact side-by-side, yielding a purely repulsive potential. In this orientation (b), the lower-order approximations are essentially indistinguishable from the full four-body energy.

\subsection{3D configurations}

\begin{figure}
    \includegraphics[width=0.9\columnwidth]{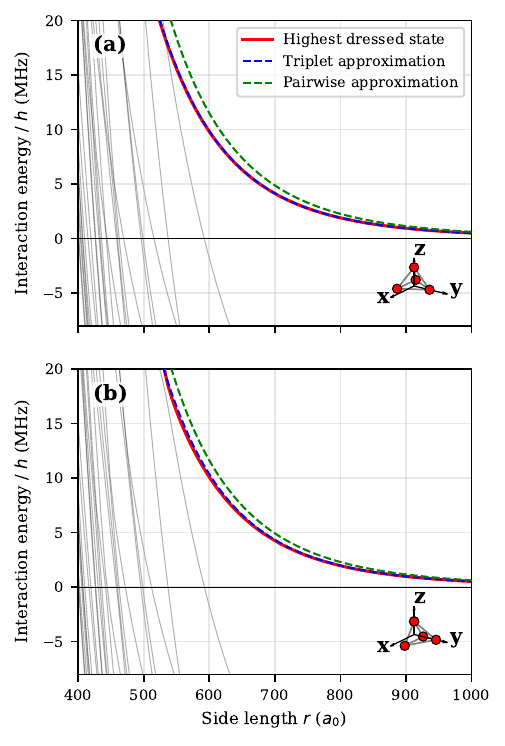}
    \caption{\label{FIG:tetrahedron} 
    Interaction energy of four dipoles arranged in a regular tetrahedron with the base oriented in the $xy$-plane (a) and $xz$-plane (b). The solid red curve represents the exact highest-energy dressed state, while the dashed green and blue curves show the pairwise and triplet approximations, respectively.
    }
\end{figure}

As an example of a three-dimensional configuration, we examine the interaction energy for a regular tetrahedron, shown in Fig.~\ref{FIG:tetrahedron}. With the base oriented in the $xy$-plane (a), the triplet approximation closely tracks the full four-body energy. The pairwise approximation deviates, which is not surprising considering the equilateral triangle in the $xy$-plane in Fig.~\ref{FIG:triangle}. The tetrahedron base would be attractive, but the repulsion due to the additional fourth molecule dominates the interaction energy. If we do not constrain the molecules to a regular tetrahedron, but apply the geometric Nelder-Mead energy minimization to an initial tetrahedral configuration, the additional molecule is moved away due to the out-of-plane repulsion. The remaining three particles adjust their intermolecular spacing to the optimal two-dimensional equilateral triangle minimum, which confirms that the tetrahedron indeed represents an energetically unfavorable configuration.

When the tetrahedron is rotated into the $xz$-plane (b), the full energy and the lower-order approximations maintain this close agreement, demonstrating that three-dimensional geometric constraints suppress the non-additivity observed in planar structures. As mentioned above, reduced symmetry leads to avoided crossing, which can be seen when the tetrahedron base is rotated from the polarization plane $xy$ to the $xz$-plane: the crossing at about $18$ MHz in (a) turns into an avoided crossing in (b).


\section{Conclusions}

We studied the effective interaction between up to four single MW-shielded molecules and found that it is not the sum of pairwise interactions, but should be treated as a many-body potential.
The most significant corrections to the pairwise approximation are for the attraction of equidistant configurations in the $xy$-plane, in particular for the depth of the potential well and the position of the shielding barrier. The attractive regime of the potential is relevant for few-body bound states and many-body bound states at high densities, such as self-bound quantum liquids and crystals. Hence, we expect that $N$-body corrections are unavoidable for accurate calculations of the resonance positions and binding energy of $N$-body field-linked states, as well as to predict their universality in the sense of Efimov state character. They should also affect the phase transition from the droplet to the self-bound quantum-crystal phase. Accurate guidance of experiments is important, as initial experimental signatures of scattering resonances and particle correlations are often based on particle loss \cite{Chen2023,zhang2026observation}, which can be masked by or confused with other loss mechanisms, such as Floquet resonances \cite{Karman2025Double}.

The corrections to the shape of the interaction potential well indicate that the avoided crossing with lower-lying dressed states shrinks. Consequently, coupling to inelastic scattering channels should increase. Predictions of the inelastic few-body collision coefficients will require a complete calculation of the corresponding dressed few-body PESs.

For the equidistant planar configurations, we found significantly larger corrections for four particles than for three. The convergence behavior of the corrections with increasing particle number is not clear yet. Given that there are six nearest neighbors in the predicted self-bound quantum-crystal membranes \cite{Ciardi2025}, the required corrections might be even greater in this scenario. This shows that studies with $N>4$ will be essential to extend our work to the many-body regime.

Although we consider here single-MW shielding with pure circular MW polarization, we expect qualitatively similar results for the effective interactions of double MW-shielded molecules. Our treatment will be extended to double MW shielding and to elliptical polarization for its relevance in BEC formation and droplet formation, respectively.

\section*{Acknowledgements}
We thank Jinglun Li, Georgios Koutentakis and Ragheed Al Hyder for discussions.
A.\ S.\ acknowledges support through the ERC grant UltraMeDiQs (project No.\ 101219560). Funded by the European Union. Views and opinions expressed are however those of the authors only and do not necessarily reflect those of the European Union or the European Research Council Executive Agency. Neither the European Union nor the granting authority can be held responsible for them.

\appendix

\section{The dipole-dipole interaction in spherical tensors}
\label{app:spherical_tensor}

In Cartesian form, the dipole-dipole interaction mixes the spatial coordinates ($\hat{r}$) with the molecular dipole operators ($\hat{\mathbf{d}}$):
\begin{equation}
    V_{dd}(\mathbf{r}) = \frac{1}{4\pi\varepsilon_0 r^3} \left[ \hat{\mathbf{d}}^{(i)} \cdot \hat{\mathbf{d}}^{(j)} - 3(\hat{\mathbf{d}}^{(i)} \cdot \hat{r})(\hat{\mathbf{d}}^{(j)} \cdot \hat{r}) \right].
\end{equation}
We can separate them by rewriting the term in brackets as the dot product of two tensors: a spatial tensor $T_{\alpha\beta} = \delta_{\alpha\beta} - 3\hat{r}_\alpha\hat{r}_\beta$ and a dipole tensor $D_{\alpha\beta}=\hat{d}^{(i)}_{\alpha}\hat{d}^{(j)}_{\beta}$, where $\alpha,\beta\,\in\{x,y,z\}$.

We split $T_{\alpha\beta}$ into its scalar (trace), antisymmetric, and traceless symmetric parts:
\begin{equation}
    \begin{split}
        T_{\alpha\beta} &= \frac{1}{3}\text{Tr}(T)\delta_{\alpha\beta} + \frac{1}{2}(T_{\alpha\beta}- T_{\beta\alpha}) \\ &\quad + \left[ \frac{1}{2}(T_{\alpha\beta} + T_{\beta\alpha}) - \frac{1}{3}\text{Tr}(T)\delta_{\alpha\beta} \right].
    \end{split}
\end{equation}
In our case, the trace vanishes ($\text{Tr}(T)=0$), as does the antisymmetric part. Therefore, $T_{\alpha\beta}$ is purely a symmetric, traceless rank-2 tensor. Because tensors of different ranks do not couple, $T_{\alpha\beta}$ only couples to the rank-2 part of the dipole tensor, $D_{\alpha\beta}$.

It is mathematically convenient to express $T_{\alpha\beta}$ and $D_{\alpha\beta}$ as spherical tensors, which transform like spherical harmonics under rotation. The interaction becomes a sum over the five rank-2 components ($p = -2, \dots, 2$):
\begin{equation}
	V_{dd} \propto \sum_{p=-2}^2 T_p^* D_p.
\end{equation}
Because $T_{\alpha\beta}$ depends only on the unit vector $\hat{r}$, its spherical components are proportional to the $l=2$ spherical harmonics. We use the Racah-normalized spherical harmonics, $C_{2,p}(\theta_r, \varphi_r) = \sqrt{\frac{4\pi}{5}} Y_{2,p}(\theta_r, \varphi_r)$.

To construct a rank-2 spherical tensor from two rank-1 vectors $\mathbf{A}$ and $\mathbf{B}$, we use the standard angular momentum addition rule with Clebsch-Gordan coefficients:
\begin{equation}
	[\mathbf{A} \otimes \mathbf{B}]_p = \sum_{q_1, q_2} \langle 1, q_1; 1, q_2 | 2, p \rangle A_{q_1} B_{q_2}.
	\label{eq:app_tensor_product}
\end{equation}
Applying this rule to our spatial unit vector $\hat{r}$ allows us to determine the exact scaling constant by evaluating the $p=0$ component. Expressing the Cartesian unit vectors in the spherical basis ($\hat{r}_0 = \hat{r}_z$, and $\hat{r}_{\pm 1} = \mp(\hat{r}_x \pm i\hat{r}_y)/\sqrt{2}$), the expansion of $[\hat{r} \otimes \hat{r}]_0$ limits non-zero terms to $q_1 + q_2 = p = 0$:
\begin{align}
	[\hat{r} \otimes \hat{r}]_0 &= \frac{1}{\sqrt{6}} (\hat{r}_1 \hat{r}_{-1}) + \frac{2}{\sqrt{6}} (\hat{r}_0 \hat{r}_0) + \frac{1}{\sqrt{6}} (\hat{r}_{-1} \hat{r}_1) \nonumber \\
	&= \frac{2}{\sqrt{6}} (\hat{r}_z^2) - \frac{2}{\sqrt{6}} \left[ \frac{1}{2}(\hat{r}_x^2 + \hat{r}_y^2) \right].
\end{align}
Using the unit vector constraint $\hat{r}_x^2 + \hat{r}_y^2 = 1 - \hat{r}_z^2$, this simplifies to
\begin{equation}
	[\hat{r} \otimes \hat{r}]_0 = \frac{1}{\sqrt{6}} (3\hat{r}_z^2 - 1).
\end{equation}
Comparing this to the $p=0$ Racah harmonic, $C_{2,0}(\theta, \varphi) = \frac{1}{2}(3\cos^2\theta - 1) = \frac{1}{2}(3\hat{r}_z^2 - 1)$, we establish the exact scaling relation $[\hat{r} \otimes \hat{r}]_0 = \frac{2}{\sqrt{6}} C_{2,0}$. Consequently, the spatial tensor becomes $T_0 = -3[\hat{r} \otimes \hat{r}]_0 = -\sqrt{6} C_{2,0}$. 

Finally, applying the Condon-Shortley phase convention $T_p^* = (-1)^p T_{-p}$, we arrive at the exact spatial-angular separated form of the dipole-dipole potential:
\begin{equation}
	V_{dd}(\mathbf{r}) = -\frac{\sqrt{6}}{4\pi\varepsilon_0 r^3} \sum_{p=-2}^{2} (-1)^p C_{2,-p}(\theta_r, \varphi_r) \left[ \hat{d}^{(i)} \otimes \hat{d}^{(j)} \right]_p.
\end{equation}
More details can be found in textbooks on angular momentum algebra \cite{zare1988}.

\section{Rotating wave approximation of the interaction}
\label{app:rwa_derivation}

To derive the analytical expression for $V_{dd}^{\text{RWA}}$ used in the main text, we adopt a formalism similar to the derivation provided in the supplementary material of Deng \textit{et al.} \cite{DengDouble}. 

By restricting the internal Hilbert space to the lowest rotational states $\{J=0, 1\}$, parity selection rules ($\Delta J = \pm 1$) dictate that the single-molecule bare dipole operator $\hat{d}_q^{(i)}$ only couples states between the ground and first excited manifolds. The dipole operator expanded in this truncated subspace is:
\begin{equation}
    \begin{split}
        \hat{d}_q^{(i)} &= \sum_{m=-1}^{1} \Big( |1,m\rangle_i \langle 1,m | \hat{d}_q | 0,0 \rangle \langle 0,0|_i \\ &\quad \;+\; |0,0\rangle_i \langle 0,0 | \hat{d}_q | 1,m \rangle \langle 1,m|_i \Big).
    \end{split}
\end{equation}

The transition matrix elements can be evaluated using standard angular momentum algebra. For absorption from the ground state, $\langle 1,m | \hat{d}_q | 0,0 \rangle = \frac{d}{\sqrt{3}} \delta_{m,q}$. For emission, the complex conjugate property of the spherical harmonics introduces a phase factor, yielding $\langle 0,0 | \hat{d}_q | 1,m \rangle = \frac{d}{\sqrt{3}} (-1)^q \delta_{m,-q}$. 

Substituting these evaluated matrix elements back into the expansion collapses the summation over $m$:
\begin{equation}
    \hat{d}_q^{(i)} = \frac{d}{\sqrt{3}} \Big( |1,q\rangle_i\langle0,0|_i + (-1)^q |0,0\rangle_i\langle1,-q|_i \Big).
\end{equation}

This allows us to express the dipole operator in this truncated space as the sum of a dimensionless raising (excitation) and lowering (de-excitation) operator:
\begin{equation}
    \hat{d}_q^{(i)} = \frac{d}{\sqrt{3}} \left( \hat{d}_q^{+(i)} + \hat{d}_q^{-(i)} \right),
\end{equation}
where $\hat{d}_q^{+(i)} = |1,q\rangle_i\langle 0,0|_i$ absorbs a MW photon, and $\hat{d}_q^{-(i)} = (-1)^q |0,0\rangle_i\langle 1,-q|_i$ emits a photon.

Applying the single-molecule rotating frame transformation $U_{\text{rot}}^{(i)}$ attaches a positive phase factor to the raising operator and a negative phase factor to the lowering operator:
\begin{equation}
    \tilde{d}_q^{(i)}(t) = U_{\text{rot}}^{(i)\dagger} \hat{d}_q^{(i)} U_{\text{rot}}^{(i)} = \frac{d}{\sqrt{3}} \left( e^{i\omega t} \hat{d}_q^{+(i)} + e^{-i\omega t} \hat{d}_q^{-(i)} \right).
\end{equation}

Substituting these time-dependent dipole operators into the rank-2 tensor $[\tilde{d}^{(i)}(t)\otimes\tilde{d}^{(j)}(t)]_{p}$ generates four distinct physical processes: pair excitation ($\sim e^{2i\omega t}$), pair de-excitation ($\sim e^{-2i\omega t}$), and resonant exchange:
\begin{equation}
    \begin{split}
        \tilde{d}_{q_1}^{(i)}(t) \tilde{d}_{q_2}^{(j)}(t) &= \frac{d^2}{3} \Big[ \hat{d}_{q_1}^{+(i)} \hat{d}_{q_2}^{+(j)} e^{2i\omega t} + \hat{d}_{q_1}^{-(i)} \hat{d}_{q_2}^{-(j)} e^{-2i\omega t} \\
        &\quad + \left( \hat{d}_{q_1}^{+(i)} \hat{d}_{q_2}^{-(j)} + \hat{d}_{q_1}^{-(i)} \hat{d}_{q_2}^{+(j)} \right) \Big].
    \end{split}
\end{equation}

The Rotating Wave Approximation (RWA) mathematically corresponds to taking the time-average of this interaction over the fast optical period $2\pi/\omega$. This explicitly integrates the highly oscillatory $e^{\pm 2i\omega t}$ terms to zero. The only surviving terms are the time-independent resonant exchange terms, enforcing the conservation of rotational excitations ($\Delta J_i + \Delta J_j = 0$). 

Inserting these surviving exchange operators back into the spatial spherical tensor yields the final analytical RWA interaction:
\begin{equation}
    \begin{split}
        V_{dd}^{\text{RWA}} &= -\frac{\sqrt{6} d^2}{12\pi\varepsilon_0 r^3} \sum_{p=-2}^{2} (-1)^p C_{2,-p}(\theta_r, \varphi_r) \\ &\quad \sum_{q_1, q_2} \langle 1, q_1; 1, q_2 | 2, p \rangle \left( \hat{d}_{q_1}^{+(i)} \hat{d}_{q_2}^{-(j)} + \hat{d}_{q_1}^{-(i)} \hat{d}_{q_2}^{+(j)} \right).
    \end{split}
\end{equation}


\vspace{86pt} 

\bibliography{references}

\end{document}